\documentstyle[12pt]{article}
\newcommand{\sect}[1]{\section{#1}\setcounter{equation}{0}}
\def\theequation{\thesection.\arabic{equation}}
\def\thesection{\arabic{section}}
\def\be{\begin{equation}}
\def\ee{\end{equation}}
\def\ben{\begin{eqnarray}}
\def\een{\end{eqnarray}}
\begin{document}


\begin{titlepage}
\pagestyle{empty}
\rightline{UMN--TH--1763/99}
\rightline{TPI--MINN--99/27}
\rightline{gr-qc/9905032}
\rightline{May 1999}
\vskip.4in
\begin{center}

{\large{\bf G\"{o}del-type Universes in String-inspired Charged Gravity}}
\end{center}
\begin{center}
\vskip 0.35in

{P. Kanti$^{\,(a)}$ and C.E. Vayonakis$^{\,(b)}$}\\[6mm]

$^{(a)}$\,{\it
{Theoretical Physics Institute, School of Physics and Astronomy, \\[1mm]
University of Minnesota, Minneapolis, MN 55455, USA}}\\[4mm]

$^{(b)}$\,{\it Division of Theoretical Physics, Physics Department,\\[1mm]
University of Ioannina, Ioannina GR-451 10, Greece}

\vskip 0.4in
{\bf Abstract}

\end{center}
\noindent
We consider a string-inspired, gravitational theory of scalar and
electromagnetic fields and we investigate the existence of axially-symmetric,
G\"{o}del-type cosmological solutions. The neutral case is studied first and
an ``extreme" G\"{o}del-type rotating solution, that respects the causality,
is determined. The charged case is considered next and two new configurations
for the, minimally-coupled to gravity, electromagnetic field are
presented. Another configuration motivated by the expected distribution
of currents and charges in a rotating universe is studied and 
shown to lead to a G\"{o}del-type solution for a space-dependent coupling
function. Finally, we investigate the existence of G\"{o}del-type cosmological
solutions in the framework of the one-loop corrected superstring effective
action and we determine the sole configuration of the electromagnetic field
that leads to such a solution. It turns out that, in all the charged cases
considered, Closed Timelike Curves do appear and the causality is always
violated.

\vspace*{30mm}
\begin{flushleft}
\begin{tabular}{l} \\ \hline
{\small Emails: yiota@physics.umn.edu, cevayona@cc.uoi.gr}
\end{tabular}
\end{flushleft}

\end{titlepage}


%
\sect{Introduction}
\paragraph{}
Einstein's Theory of Relativity has been exhaustively studied for many
decades and a plethora of solutions, both mathematically and physically
interesting, have been determined. One of the most fascinating
cosmological solutions that emerged from Einstein's equations was the
G\"{o}del-type rotating universe~\cite{godel}. The feature that was most 
appealing and intensely attracted the interest of scientists to these
cosmological models was the fact that they admit Closed Timelike Curves.
This leads to the violation of causality in the universe~\cite{hawk}
and to the possibility of time-travel into our past or future~\cite{weyl}. 

Since their discovery in 1949, several versions and modifications of the
G\"{o}del cosmological solutions have been considered. Apart from the perfect
fluid that gave rise to the original G\"{o}del solutions, extra fields,
i.e. vector~\cite{som}~\cite{reboucas}~\cite{ray}, scalar~\cite{scalar},
spinor~\cite{spinor} (and their
combinations~\cite{comb1}~\cite{reb}~\cite{comb2}) and even
tachyon fields~\cite{tachyon}, were gradually introduced. Higher-derivative
gravitational terms were also considered in an attempt to introduce
quantum corrections to Einstein's theory of gravity~\cite{matsas}.
Recently, geometrical aspects of G\"{o}del-type solutions have also been
studied, either determining solutions that could be interpreted as the
exterior of G\"{o}del spacetimes~\cite{mac} or connecting the 3+1 
G\"{o}del geometry
with a 2+1 anti-de Sitter subspace~\cite{desitter}. Five-dimensional
models that admit generalized G\"{o}del-type solutions have also been
constructed~\cite{5D} and the same type of solutions was proved to arise 
also in Riemann-Cartan spacetimes with non-zero torsion~\cite{cartan}.  

As it is well known, superstring theory is the most plausible theory
nowadays for the unification of all forces and the formulation of a
quantum theory of gravity. The effective theory of heterotic superstrings
at low energies provides us with a gravitational theory from which the
Einstein's theory of gravitation automatically emerges when one ignores
all higher-loop stringy corrections and assigns constant values to the scalar
fields of the theory. On the other hand, when one takes into account these
extra terms, a generalized theory of gravity arises which leads to 
important modifications when compared to the Theory of General Relativity. 
The question of what happens to the causal pathologies of the G\"{o}del-type
cosmological solutions when one introduces stringy corrections to Einstein's
theory of gravity inevitably arises. It is, then, necessary to investigate
the existence of G\"{o}del-type solutions and the appearance or not of Closed
Timelike Curves in the framework of the superstring effective theory. Barrow
and Dabrowski~\cite{barrow} tried to answer this question and studied
the one-loop corrected superstring effective theory in the string frame
(and the lowest order in the Einstein frame) by considering
a non-zero dilaton and axion field. They found that, in the context of this
specific theory, G\"{o}del universes do not contain Closed Timelike Curves
and the causality need not be violated. Here, we will try to extend their
work and study the full one-loop superstring effective theory in the
Einstein frame. We will consider all four scalar fields, as well as
higher-derivative gravitational and electromagnetic terms. However, our
analysis will be greatly simplified by the fact that all higher-derivative
gravitational terms identically vanish for a G\"{o}del spacetime.

The structure of this article is as follows~: in section 2, we present
the action functional of the theory and derive the corresponding equations
of motion in an axially-symmetric spacetime background. In section 3, we
study the neutral case where scalar fields minimally-coupled to gravity give
rise to G\"{o}del-type cosmological solutions. The charged case is considered
in section 4 and the equations of motion are conveniently formulated. 
In the same section, we deal with the problem of the simultaneous existence
of an electromagnetic and scalar fields all minimally-coupled to gravity.
All the possible configurations for the electromagnetic components that
lead to G\"{o}del-type rotating universes are determined and the solution for
the scalar fields is presented. In section 5, the physically favoured case
of a constant magnetic field along the rotation axis together with a radial
electric and magnetic component is studied and proved to be compatible with
an axially-symmetric universe only in the case of a space-dependent coupling
function. The non-trivial interaction 
between the electromagnetic and the scalar fields in the framework of 
the one-loop corrected, superstring, effective theory is the basic feature
of the gravitational theory considered in section 6. There, the possible
configurations of the electromagnetic field that lead to G\"{o}del-type
universes are re-examined and the solution for the scalar fields is
determined. Section 7 is devoted to the conclusions derived from our
analysis while some useful formulae and equations are displayed in
the Appendix at the end of the paper.


%
\sect{Action Functional and Equations of Motion of the Theory}
\paragraph{}
We consider the following, string-inspired, generalized theory of gravity
which describes the coupling of four scalar fields and the electromagnetic
field to gravity
\vspace*{3mm}
%
\ben
S_{eff}&=&\int d^4x\,\sqrt{-g}\,\Biggl\{\frac{R}{2}  
+ \frac{1}{4}\,(\partial_\mu \phi)^2 
+ \frac{1}{4}\,e^{-2 \phi}\,(\partial_\mu a)^2
+ \frac{3}{4}\,(\partial_\mu\sigma)^2 
\nonumber\\[3mm] &+& 
\frac{3}{4}\,e^{-2\sigma}\,(\partial_\mu b)^2+
f_1\,{\cal{R}}^2 _{GB} 
+  f_2\,{\cal{R}} {\tilde{\cal{R}}} + f_3\,F^{\mu \nu} F_{\mu \nu}
+ f_4\,F {\tilde{F}}\Biggr\}\,\,.
\label{final}
\een\\  
In the above, $\phi$ is the dilaton, $a$ and $b$ the axions and $\sigma$ the 
modulus field. The gravitational and electromagnetic quantities that appear
in the above expression are defined as
\ben
{\cal R}^2_{GB} & \equiv & R_{\mu\nu\kappa\lambda}\,R^{\mu\nu\kappa\lambda}
-4\,R_{\mu\nu}\,R^{\mu\nu} + R^2\,\,, \\[3mm]
{\cal R}\tilde{\cal R} & \equiv& \eta^{\mu\nu\rho\sigma}\,R^{\kappa\lambda}
_{\,\,\,\,\mu\nu}\,R_{\rho\sigma\kappa\lambda}\,\,,\\[3mm]
F \tilde{F} &\equiv& \eta^{\mu\nu\rho\sigma}\,F_{\mu\nu}\,
F_{\rho\sigma}\,\,.
\een

In the framework of the one-loop corrected, effective theory of heterotic
superstrings at low energies the coupling functions $f_i$ have the form
\ben
 f_1 \equiv \alpha'\Biggl({\frac {e^{\phi}} {8 g^2}} + \Delta\Biggr)
\quad &,& \quad  
f_2 \equiv \alpha'\Biggl({\frac {a} {8 g^2}} + \Theta\Biggr)\,\,, 
\nonumber \\[4mm] 
 f_3 \equiv \alpha'\Biggl({-\frac {e^{\phi}} {8 g^2}} + \hat{\Delta}\Biggr)
 \quad &,& \quad
f_4 \equiv \alpha'\Biggl({-\frac {a} {8 g^2}} + \hat{\Theta}\Biggr)\,\,,
\label{string}
\een
where $\alpha'$ is the inverse string tension and
the functions $\Delta$, $\Theta$, $\hat{\Delta}$ and $\hat{\Theta}$
depend only on the scalar fields $b$ and $\sigma$. However, during the
following sections, we are going to assume that the coupling functions
$f_i$ are zero, constant, space-dependent or field-dependent according
to the type of gravitational theory considered each time.
\paragraph{}
We assume that the universe is described by the following axially-symmetric
line-element
\ben
ds^2 &=& [\,dt+C(r)\,d\varphi\,]^2-dr^2-dz^2-D^2(r)\,d\varphi^2
\nonumber \\[3mm]
     &=& dt^2-dr^2-dz^2-(D^2-C^2)\,d\varphi^2 + 2C(r)\,dt\,d\varphi\,\,.
\label{metric}
\een
By making use of the non-vanishing
components of the curvature tensor $R_{\mu\nu\rho\sigma}$,
of the Ricci tensor $R_{\mu\nu}$ and the scalar curvature $R$, which
are given in the Appendix, it is easy to see that both of the
higher-derivative gravitational terms ${\cal R}^2_{GB}$ and
${\cal R}\tilde{\cal R}$ vanish identically. Then,
the variation of the action functional of the theory with respect to the
scalar fields, the metric tensor and the electromagnetic potential leads to
the following set of equations of motion
%
\vspace*{1mm}
%
\be
\frac {1}{\sqrt{-g}}\,\partial_{\mu}[\sqrt{-g}\,\partial^{\mu} \phi]
 = -e^{-2 \phi} (\partial_{\mu}a)^2 + 2\,\frac{\partial f_3}{\partial\phi}
\,F^{\mu \nu} F_{\mu \nu}\,\,,
\label{dilaton}
\ee
%
\vspace*{2mm}
%
\begin{equation}
\frac{1}{\sqrt{-g}}\, \partial_{\mu}[\sqrt{-g} e^{-2 \phi} 
\,\partial^{\mu} a]
= 2\,\frac {\partial f_4}{\partial a}\,F {\tilde{F}}\,\,,
\label{axion1}
\end{equation}
%
\vspace*{2mm}
%
\ben
\frac{1}{\sqrt{-g}} \partial_{\mu}[\sqrt{-g} \partial^{\mu} \sigma] 
=-e^{-2 \sigma}{({\partial} _{\mu} b)}^2 +  \frac{2}{3}\,
\frac{\partial f_3}{\partial\sigma}\,F^{\mu\nu}  F_{\mu \nu}  +
\frac{2}{3}\,\frac{\partial f_4}{\partial\sigma}\,F {\tilde{F}}\,\,,
\label{modulus}     
\een     
%
\vspace*{2mm}
%
\ben
\frac{1}{\sqrt{-g}} \partial_{\mu}[\sqrt{-g} e^{-2 \sigma}
\partial^{\mu}b] =  \frac{2}{3}\,\frac{\partial f_3}{\partial b}\,
F^{\mu\nu} F_{\mu \nu}  +
\frac{2}{3}\,\frac{\partial f_4}{\partial b}\,F {\tilde{F}}\,\,,
\label{axion2}         
\een
\vspace*{3mm}
\begin{eqnarray}
&~& R_{\mu\nu}-\frac{1}{2}\,g_{\mu \nu}R 
+ 4\,f_3\,(F_{\mu} ^{\;\;\sigma} F_{\nu \sigma}-\frac{1}{4}
\,g_{\mu \nu} F^{\rho\sigma} F_{\rho \sigma}) =
-\frac{1}{2}\,(\partial_\mu\phi)\,(\partial_{\nu} \phi)
\nonumber \\[3mm]
&~&+\,\frac{1}{4}\,g_{\mu\nu}\,(\partial_{\rho}\phi)^2
-\frac{e^{-2 \phi}}{2}\,(\partial_\mu a)\,(\partial_{\nu}a) 
+\frac{1}{4}\,g_{\mu \nu} e^{-2 \phi}(\partial_{\rho}a)^2
-\frac{3}{2}\,(\partial_\mu \sigma)\,(\partial_\nu \sigma)
\label{gravity} \\[3mm]
&~&+ \,\frac{3}{4}\,g_{\mu\nu}(\partial_{\rho}\sigma)^2 
-\frac {3}{2}\,e^{-2\sigma}(\partial_{\mu}b)\,(\partial_{\nu}b) 
+\frac {3}{4}\,e^{-2\sigma} g_{\mu \nu }
(\partial_{\rho}b)^2\,\,,
\nonumber   
\end{eqnarray} 
%
\vspace*{3mm}
%
\begin{equation}
\hspace*{-1.5cm}
\frac{1}{\sqrt{-g}}\,\partial_\mu[\,\sqrt{-g}\,f_3\,F^{\mu \nu}\,] + 
\frac{1}{\sqrt{-g}}\,\partial_\mu[\,\sqrt{-g}\,f_4\,{\tilde{F}}
^{\mu\nu}\,]=0\,\,,
\label{abelian}
\end{equation}\\
where we have assumed that the functions $f_3$ and $f_4$ depend
on spacetime either directly of through their dependence on the scalar
fields. In the latter case, there is a non-trivial coupling between the
scalar fields of the theory and the electromagnetic field.


%
\sect{ The Neutral Case}
\paragraph{}
In this section, we ignore the electromagnetic field and concentrate
our attention on the Einstein's equations, $G_{\mu\nu}=-T_{\mu\nu}^s$
where $T_{\mu\nu}^s$ is the energy-momentum tensor of the scalar fields.
The non-vanishing components of both the Einstein and the 
energy-momentum tensor are also given in the Appendix. We notice that
simple relations hold between the various components of $T_{\mu\nu}^s$,
namely
\be
T_{zz}^s=-T_{rr}^s \quad , \quad T_{\varphi\varphi}^s=(C^2-D^2)\,T_{tt}^s 
\quad , \quad T_{t\varphi}^s=C\,T_{tt}^s\,\,. \label{relations}
\ee
By making use of the above relations, we can determine the solution
for the metric functions $C$ and $D$ in the presence of multiple scalar
fields in the theory. Rearranging the ($tt$) and ($t\varphi$) components
of the Einstein's equation as well as the ($tt$) and ($\varphi\varphi$)
components, we find the constraints
\be
C''=\frac{C'D'}{D} \qquad , \qquad {C'}^2=D D''\,\,,
\label{con1}
\ee
respectively, where the prime denotes derivative with respect to the
radial coordinate $r$. However, doing the same with the ($rr$) and
($zz$) components,
we are led to the result $D''=0$. Then, the general solution of the above
system is~: $D(r)=d_1\,r+d_2$ and $C(r)=c_1$. If the line-element
(\ref{metric}) is to describe a rotating universe with the function $C(r)$
being the analog of its angular momentum, then on the
$z$-axis, that is in the limit $r \rightarrow 0$, both of the functions
$C(r)$ and $D^2-C^2$ must vanish. This leads to $c_1=d_2=0$ and the 
vanishing of the $g_{t\varphi}$ component of the metric tensor. Redefining
the angular coordinate $\varphi$ in order to absorb the arbitrary constant
$d_1$, the line-element (\ref{metric}) reduces to that of a non-rotating,
spherically-symmetric universe. Finally, by rearranging the $(tt)$ and 
$(rr)$ components of the Einstein's equations, we obtain the following
relations
\ben
&~& (\partial_r \phi)^2 + e^{-2\phi}\,(\partial_r a)^2 +
3\,(\partial_r \sigma)^2 + 3\,e^{-2\sigma}\,(\partial_r b)^2=0\,\,, \\[4mm]
&~& (\partial_z \phi)^2 + e^{-2\phi}\,(\partial_z a)^2 +
3\,(\partial_z \sigma)^2 + 3\,e^{-2\sigma}\,(\partial_z b)^2=0\,\,.
\een
According to the above constraints, each one of the derivatives of the scalar
fields with respect to $r$ or $z$ must vanish. In other words, the only
acceptable solution for the scalar fields is the trivial one~:
$\phi=\phi_0$, $a=a_0$, $\sigma=\sigma_0$ and $b=b_0$ where $\phi_0$,
$a_0$, $\sigma_0$ and $b_0$ are arbitrary constants.
\paragraph{}
However, if we include in the theory the cosmological constant $\Lambda$,
the Einstein tensor $G_{\mu\nu}$ should be replaced by $G_{\mu\nu}+
\Lambda\,g_{\mu\nu}$. In this case, the constraints (\ref{con1}) are
still valid but, now, the rearrangement of the ($rr$) and ($zz$) components
of the Einstein's equations gives the result
\be
\frac{D''}{D}=-2\Lambda\,\,.
\label{con2}
\ee
The combination of the above relation with the second one of (\ref{con1})
leads to~: ${C'}^2=-2\Lambda\,D^2$ which means that the cosmological constant
must be always negative in order to ensure the reality of the functions
$C$ and $D$. If we demand that, in the limit $r \rightarrow 0$, both of the
functions $C$ and $D^2-C^2$ must vanish, the general solution of the system
(\ref{con1})-(\ref{con2}) takes the form
\ben
&~& D(r)=\frac{1}{m}\,sinh(mr)\,\,, \label{causald}\\[3mm]
&~& C(r)=\frac{1}{m}\,[\,cosh(mr) -1\,]=
\frac{2}{m}\,sinh^2(\frac{mr}{2})\,\,,
\label{causalc}
\een
where $m^2=-2\Lambda$. The above solution for the spacetime is a G\"{o}del-type
solution which is defined as~\cite{godel}~\cite{reb}
\be
ds^2=\biggl[dt+\frac{4\Omega}{m^2}\,sinh^2\bigl(\frac{mr}{2}\bigr)
\,d\varphi\biggr]^2 - dr^2-dz^2-\frac{sinh^2(mr)}{m^2}\,d\varphi^2\,\,,
\label{godel}
\ee
where again $m^2=-2\Lambda$ and $\Omega$ is the rate of the rigid rotation
of the matter around the $z$-axis. In the case of the original G\"{o}del
universe, the matter is a perfect-fluid with an energy-momentum tensor of the
form $T_{\mu\nu}=\rho\,V_\mu V_\nu$, where $V^\mu=\delta^\mu_0$ the fluid
four-vector velocity and $\rho$ the density of matter. 
The $g_{\varphi\varphi}$ component of the G\"{o}del metric (\ref{godel}) has
the form
\be
g_{\varphi\varphi}=-\frac{4}{m^2}\,sinh^2(\frac{mr}{2})\,\biggl[1+
\biggl(1-\frac{4\Omega^2}{m^2}\biggr)\,sinh^2\bigl(\frac{mr}{2}\bigr)\biggr]
\label{ctcs}
\ee
and it is obvious that for $4\Omega^2>m^2$, the $g_{\varphi\varphi}$ component
may become positive for a range of values of the radial coordinate $r$
leading to the existence of Closed Timelike Curves (CTC's). The solution found
above (\ref{causald})-(\ref{causalc}) constitutes a G\"{o}del-type solution
with $4\Omega^2=m^2$. It is, actually, a limiting case of a G\"{o}del-type
universe that ensures the non-existence of Closed Timelike Curves.
The same type of solution was found in the case of an electromagnetic
field minimally coupled to gravity~\cite{reb} and in the framework of
a higher-derivative gravi\-tational theory~\cite{matsas} and it was shown       
that this ``extreme" G\"{o}del-type solution, although ``on the verge"
of displaying the breakdown of causality, is free of any Closed Timelike
Curves~\cite{matsas}~\cite{calvao}.

Next, we concentrate our attention on the scalar fields. Their dependence
on the spatial coordinates is strongly restricted by the constraint
\be
\partial_r \phi\,\partial_z \phi+ e^{-2\phi}\,\partial_r a\,\partial_z a
+3\,\partial_r \sigma\,\partial_z \sigma+
3e^{-2\sigma}\,\partial_r b\,\partial_z b=0\,\,,
\label{con3}
\ee
which follows from the ($rz$) component of the Einstein's equations. We shall
consider the most general case in which all the scalar fields depend 
on the $r$ and $z$ coordinates due to the staticity and axial symmetry of
the spacetime. The remaining components of the Einstein's equations give
the following constraints
\ben
-2\Lambda &=& (\partial_r \phi)^2 + (\partial_z \phi)^2 +
e^{-2\phi}\,[\,(\partial_r a)^2 + (\partial_z a)^2\,] +
3\,[\,(\partial_r \sigma)^2 + (\partial_z \sigma)^2\,]\nonumber\\[3mm]
&+& 3e^{-2\sigma}\,[\,(\partial_r b)^2 + (\partial_z b)^2\,]\,\,,
\label{con4}\\[6mm]
-2\Lambda &=& -(\partial_r \phi)^2 + (\partial_z \phi)^2 +
e^{-2\phi}\,[\,-(\partial_r a)^2 + (\partial_z a)^2\,] +
3\,[\,-(\partial_r \sigma)^2 + (\partial_z \sigma)^2\,]\nonumber\\[3mm]
&+& 3e^{-2\sigma}\,[\,-(\partial_r b)^2 + (\partial_z b)^2\,]\,\,.
\label{con5}
\een
Rearranging the above expressions, we are led to the results
\ben
&~& (\partial_r \phi)^2 + e^{-2\phi}\,(\partial_r a)^2 +
3\,(\partial_r \sigma)^2 + 3\,e^{-2\sigma}\,(\partial_r b)^2=0\,\,,\\[4mm]
&~& (\partial_z \phi)^2 + e^{-2\phi}\,(\partial_z a)^2 +
3\,(\partial_z \sigma)^2 + 3\,e^{-2\sigma}\,(\partial_z b)^2=-2\Lambda\,\,.
\een
The former constraint is satisfied only if we assume that the scalar fields
do not depend on the radial coordinate. In this case, the supplementary
constraint (\ref{con3}) is automatically satisfied. The latter constraint
leads to the following dependence of the scalar fields on the coordinate $z$
\ben
&~& \phi=\phi_0 + \phi_1\,z \qquad , \qquad a=a_0 + a_1\,e^\phi\,\,,\\[4mm]
&~& \sigma=\sigma_0 + \sigma_1\,z \qquad , \qquad b=b_0 + b_1\,e^\sigma\,\,,
\een
where $\phi_i$, $a_i$, $\sigma_i$ and $b_i$ are constants.
If we substitute the above expressions for the scalar fields in their
equations of motion, we find some further constraints, namely $\phi_1\,a_1=0$
and $\sigma_1\,b_1=0$, which means that not both of the pairs $(\phi,
\sigma)$ and $(a, b)$ can have a non-trivial solution. If we choose to set
$\phi_1=\sigma_1=0$ in order to satisfy the above constraints, then, the
second pair of scalar fields $(a, b)$ is inevitably led to the trivial
solution as well. If, on the other hand, we set $a_1=b_1=0$, the dilaton and
modulus field can still preserve a non-trivial form. As a result, the most
interesting solution for the scalar fields is the following
\be
\phi=\phi_0+\phi_1\,z \quad , \quad a=a_0 \quad , \quad \label{solpa0}
\sigma=\sigma_0+\sigma_1\,z \quad , \quad b=b_0\,\,,
\ee
where 
\be
-2\Lambda=\phi_1^2 + 3\,\sigma_1^2\,\,.
\label{fund0}
\ee

Similar results for the dilaton and axion field $a$ were found by Barrow and
Dabrowski~\cite{barrow}. Here, we have also included the axion field $b$
and the modulus $\sigma$, that are present in the low-energy effective string
theory, and shown that the solution for the metric remains unchanged.
Moreover, from the above analysis, it is easy to conclude that a theory with
an arbitrary number of multiple scalar fields formulated in the following way
\be
S=\int d^4x\,\sqrt{-g}\,\left\{\frac{R}{2} - \Lambda  
+ \sum_{i=1}^N \Biggl(\frac{1}{4}\,\partial_\mu \phi_i\,
\partial^\mu\phi_i + \frac{1}{4}\,e^{-2 \phi_i}\,
\partial_\mu a_i\,\partial^\mu a_i\Biggr) \right\}
\ee
always accepts axially-symmetric spacetime backgrounds of the form
(\ref{metric}) and (\ref{causald})-(\ref{causalc}) with the scalar
fields given by
\be
\phi_i=\phi^i_0+\phi^i_1\,z \qquad , \qquad a=a^i_0
\ee
and with the constants $\phi^i_1$ satisfying the constraint
\be
-2\Lambda=\sum_{i=1}^N (\phi^i_1)^2 \,\,.
\ee
%


\sect{EM and Scalar Fields~: Minimal Coupling}
\paragraph{}

Next, we assume that an electromagnetic field, in the form of $F_{\mu\nu}
F^{\mu\nu}$ and $F \tilde{F}$, is included in the theory with coupling
functions $f_3$ and $f_4$, respectively. In this case, the Einstein's
equations are supplemented by Maxwell's equations that come from the
equation of motion of the electromagnetic field (\ref{abelian}). The 
exact form of both sets of equations are displayed in the Appendix.
The relations (\ref{relations}) between the components of $T_{\mu\nu}^s$
still hold and may be used once again in order to derive a number of
constraints on the metric functions $C$ and $D$. Rearranging accordingly 
eqs.(\ref{geq2})-(\ref{geq3}), (\ref{geq1})-(\ref{geq4}) and
(\ref{geq1})-(\ref{geq5}), we are led to the following constraints
\be
\frac{D''}{D}+ 2\Lambda + 4 f_3\,\Bigl[F^2_{rz} +
F^2_{t\varphi}\,(g^{t\varphi} g^{t\varphi}-
g^{tt} g^{\varphi\varphi})\Bigr]=0\,\,,
\label{gcon1}
\ee
\vspace*{4mm}
\be
C\,C''-D\,D'' + C'^2-\frac{C C' D'}{D} +
4 f_3\,\Bigl[F^2_{\varphi r} + F^2_{\varphi z}-
(C^2-D^2)\,(F_{tr}^2 + F^2_{tz})\Bigr]=0\,\,,
\label{gcon2}
\ee
\vspace*{4mm}
\be
\frac{C''}{2} -\frac{C'\,D'}{2D} + 4 f_3 \,\Bigl[F_{tr} F_{\varphi r} 
+ F_{tz} F_{\varphi z}-C\,(F^2_{tr}+F^2_{tz})\Bigl]=0\,\,. \label{gcon3}
\ee

Moreover, one could express the derivatives of the scalar fields with
respect only to $r$ or only to $z$ in terms of the metric functions and
the various components of the electromagnetic field. Adding or
subtracting the ($tt$) and ($rr$) components of the Einstein's equations and
by making use of the above constraints (\ref{gcon1})-(\ref{gcon3}),
we obtain the results
\ben
&~& \hspace*{-1cm}(\partial_r \phi)^2 + e^{-2\phi}\,(\partial_r a)^2 +
3\,(\partial_r \sigma)^2 + 3\,e^{-2\sigma}\,(\partial_r b)^2=
8 f_3 \,\Bigl(F^2_{rz}-F^2_{tr}\Bigr) \nonumber\\[3mm]
&~& \hspace*{5cm} +\,\frac{8 f_3}{D^2}\,\Bigl(F^2_{\varphi t} -
F^2_{\varphi z} -
C^2\,F^2_{tz} +2C\,F_{tz} F_{\varphi z}\Bigr)\,\,,
\label{gres1}\\[7mm]
&~& \hspace*{-1cm} (\partial_z \phi)^2 + e^{-2\phi}\,(\partial_z a)^2 +
3\,(\partial_z \sigma)^2 + 3\,e^{-2\sigma}\,(\partial_z b)^2=
\frac{C'^2}{D^2}+ 8 f_3\,\Bigl(F^2_{rz}+F^2_{tr}\Bigr)
\nonumber\\[3mm] &~& \hspace*{5cm}
+\,\frac{8 f_3}{D^2}\,\Bigl(F^2_{\varphi t} + F^2_{\varphi z} +
C^2\,F^2_{tz} - 2C\,F_{tz} F_{\varphi z}\Bigr)\,\,.
\label{gres2}
\een
The substitution of eqs.~(\ref{geq1})-(\ref{geq5}) by the constraints
(\ref{gcon1})-(\ref{gres2}) allows us to determine first the form of the
metric functions $C$ and $D$ in terms only of the non-vanishing components
of the electromagnetic field, through eqs.~(\ref{gcon1})-(\ref{gcon3}), and
subsequently the solution for the scalar fields, through
eqs.~(\ref{gres1})-(\ref{gres2}).
\paragraph{}
We start the study of the charged case by first considering the minimal
coupling of the electromagnetic and scalar fields to the gravitational field.
As a result, we are going to assume that both of the coupling functions
$f_3$ and $f_4$ do not depend on spacetime and set $f_3=f_4=-1/4$.
The necessary and sufficient conditions for the existence and space-time
homogeneity, at the same time, of a G\"{o}del-type axially-symmetric universe
are the following~\cite{reb}
\be
\frac{D''}{D}=const.=-2\Lambda=m^2 \qquad , \qquad
\frac{C'}{D}=const.= 2\Omega
\label{cons}
\ee
By studying eq.~(\ref{gcon1}), it is straightforward to see that the
first condition is satisfied if we assume that $F_{rz}=F_{t\varphi}=0$,
i.e. if that both the electric and magnetic components along the
$\hat{\varphi}$ direction vanish, an assumption easily accepted given the 
cylindrical symmetry of the problem. The second condition is satisfied
if we assume that the combination of the components of the electromagnetic
field that appears inside the brackets in eq.~(\ref{gcon3}) vanishes, too.
Then, eq.~(\ref{gcon2}) provides us with a relation between the different
parameters of the theory.  Note, that once these two conditions (\ref{cons})
are satisfied, a G\"{o}del-type cosmological solution appears
independently of the existence or not of scalar fields in our theory. 

G\"{o}del-type cosmological solutions, in the case of a minimally coupled
electromagnetic field but in the absence of scalar fields, already exist
in the literature~: the Som-Raychaudhuri solution~\cite{som} and the
Rebou\c{c}as solution~\cite{reboucas}. The Som-Raychaudhuri ansatz for the
electromagnetic field has the form
\be F^{\varphi r}(r, z)= -\frac{\sqrt{4\Omega^2-m^2}}{D(r)} \qquad ,
\qquad F^{rt}(r, z)=C(r)\,F^{\varphi r}\,\,.
\label{som}
\ee
When written in terms of covariant components, the above ansatz gives
rise only to a magnetic field along the $z$ axis~:
$B_z=F_{\varphi r}=\sqrt{4\Omega^2-m^2}\,D(r)$. In this case,
both of the conditions (\ref{cons}) are satisfied giving rise to a
G\"{o}del-type solution for the spacetime while the equations for the
electromagnetic field (\ref{abeq1})-(\ref{abeq4}) are satisfied only if we
assume the pre\-sence of a continuous electric-charge distribution with
charge density given by~: $\rho= \frac{\Omega}{2}\,\sqrt{4\Omega^2-2m^2}$.
On the other hand, the Rebou\c{c}as solution assumes the ansatz
\be
F_{tz}= A_0\,sin(2\Omega z) \quad ,
\quad F_{\varphi z}=C(r)\,F_{tz} \quad , \quad
F_{\varphi r}=-A_0\,D(r)\,cos(2\Omega z)\,\,,
\label{reb}
\ee
where $A_0=\sqrt{4\Omega^2-m^2}$. The above ansatz also satisfies the
conditions (\ref{cons}) as well as the source-free Maxwell equations
(\ref{abeq1})-(\ref{abeq4}). 

As Raychaudhuri and Guha Thakurta have pointed out in their work~\cite{ray},
there are more than one ansatzes for the form of the electromagnetic field
that may lead to the same G\"{o}del-type cosmological solution. In this
section, we make a thorough analysis of the possible configurations of the
electromagnetic field and present the remaining solutions. Since the first
of the conditions (\ref{cons}) is easily satisfied by setting
$F_{rz}=F_{t\varphi}=0$, we concentrate our attention on the constraints
(\ref{gcon2})-(\ref{gcon3}) looking for all the possible, non-trivial
combinations of the components of the electromagnetic field that respect
the second condition. We notice that these constraints are characterized
by a ``duality", i.e. they remain invariant under the interchange of the
indices $r$ and $z$ in the components of the electromagnetic field. As a
result, we are led to consider the ``dual" forms of the aforementioned
solutions (\ref{som}) and (\ref{reb}). We start with the 
Som-Raychaudhuri-dual solution and consider the ansatz
\be
B_r=F_{z\varphi}=\sqrt{4\Omega^2-m^2}\,D(r)\,\,.
\label{somdual}
\ee
It is easy to see that this ansatz satisfies both of the constraints
(\ref{cons}) leading to a G\"{o}del-type solution of the form (\ref{godel}).
As in the original Som-Raychaudhuri solution, we are forced to introduce
a source term in order to satisfy Maxwell's equations
(\ref{abeq1})-(\ref{abeq4}). However, in this case the source term
corresponds to a distribution of magnetic monopoles instead of electric.
Following Dirac's example~\cite{dirac}, we introduce, apart from the electric
four-current $j^\mu=\{\rho,\vec{j}\}$, a magnetic four-current
$k^\mu=\{\eta,\vec{k}\}$ and write Maxwell's equations
(\ref{abelian}) in the more general form
\be
\frac{1}{\sqrt{-g}}\,\partial_\mu\,\biggl[\sqrt{-g}\,(f_3\,F^{\mu\nu} +
f_4\,\tilde{F}^{\mu\nu})\biggr]=-(j^\mu+k^\mu)\,\,.
\label{max}
\ee
Then, the substitution of the ansatz (\ref{somdual}) into the above 
equation leads to the determination of a magnetic charge density equal
to~:
\be
\eta= -\frac{\sqrt{4\Omega^2-m^2}}{4}\,\frac{D'}{D}\,\,.
\ee
The magnetic monopoles are spread all over the universe as their
corresponding density vanishes at the origin while it approaches a
constant value at infinity. 

Finally, the ``dual" ansatz of the Rebou\c{c}as solution (\ref{reb}) can be
written in the form
\be
F_{tr}= \frac{\sqrt{4\Omega^2-m^2}}{\sqrt{2}} \quad , \quad
F_{\varphi r}=C(r)\,F_{tr}
\quad , \quad F_{\varphi z}= D(r)\,F_{tr}\,\,.
\label{rebdual}
\ee
As the original Rebou\c{c}as solution, the above ``dual" solution leads to
a G\"{o}del-type universe (\ref{godel}) and satisfies the source-free
Maxwell's equations. As a matter of fact, the substitution of the
above ansatz into the generalized Maxwell's equations (\ref{max})
leads to non-vanishing expressions for the electric and magnetic-charge
densities, i.e.
\be
\rho=-\frac{\sqrt{4\Omega^2-m^2}}{\sqrt{2}}\,D' \qquad , \qquad
\eta=\frac{\sqrt{4\Omega^2-m^2}}{\sqrt{2}}\,D'
\ee
which, however, being exactly opposite, cancel each other. If we set
$F_{\varphi z}=0$, the solution for the spacetime remains unchanged and,
as a result, the following combination of $F_{tr}$ and
$F_{\varphi r}$ alone
\be
F_{tr}= \sqrt{4\Omega^2-m^2} \quad , \quad F_{\varphi r}=C(r)\,F_{tr}
\label{trial}
\ee
is an acceptable solution of eqs.~(\ref{gcon1})-(\ref{gcon3}) by itself.
However, the vanishing of the component $F_{\varphi z}$ necessitates the
introduction of a source-term into Maxwell's equations. The corresponding
electric-charge density that follows has the form
\be
\rho= \frac{\sqrt{4\Omega^2-m^2}}{4}\,\frac{D'}{D}\,\,.
\ee
As in the case of the Som-Raychaudhuri-dual solution, the electric monopoles
are spread all over the universe. The ``dual" solution of this last
combination can be written as
\be
F_{tz}= \sqrt{4\Omega^2-m^2} \quad , \quad F_{\varphi z}=C(r)\,F_{tz}
\label{trialdual}
\ee
and leads, as we expect, to a G\"{o}del-type universe of the form (\ref{godel}).
The electric-charge density found above is substituted, in this dual case,
by a magnetic-charge density of the form $\eta= \frac{\Omega}{2}\,
\sqrt{4\Omega^2-m^2}$. The above density corresponds to a homogeneous,
constant distribution of magnetic monopoles.

The modification (\ref{max}) of Maxwell's equations is acceptable only
in the case where the corresponding source-terms do not act as an
additional source for the gravitational field and, thus, do not destroy
the G\"{o}del-type cosmological solutions. In other words, the introduction
of charge and current densities in the Maxwell's equations result from
the introduction of the term $-2(j^\mu+k^\mu)A_\mu$ in the Lagrangian
of the theory, where $A_\mu$ is the electromagnetic potential. In
principle, such a term could enter the Einstein's equations and cause
the violation of the G\"{o}del conditions (\ref{cons}). In the cases of 
the Som-Raychaudhuri ansatz (\ref{som}), the Rebou\c{c}as ansatz (\ref{reb})
and their ``dual" ansatzes (\ref{somdual}) and (\ref{rebdual}), the 
source-term $(j^\mu+k^\mu)A_\mu$ vanishes identically and, thus, does
not affect the Einstein's equations and the solution for the spacetime.
On the other hand, the ansatzes (\ref{trial}) and (\ref{trialdual})
introduce a non-vanishing source-term in the Lagrangian which depends
non-trivially on the spacetime coordinates and thus modifies the
G\"{o}del-type solution. As a result, these two ansatzes fail to lead to a
G\"{o}del-type rotating universe and, thus, they will not be considered
any longer.
\paragraph{}
All of the above solutions for the spacetime and electromagnetic field
are compatible with the existence of scalar fields, minimally-coupled
to gravity, in the theory. The constraints (\ref{gres1})-(\ref{gres2})
can, now, be written as
\ben
&~& \hspace*{-1.5cm}(\partial_r \phi)^2 + e^{-2\phi}\,(\partial_r a)^2 +
3\,(\partial_r \sigma)^2 + 3\,e^{-2\sigma}\,(\partial_r b)^2=
2\,\Bigl[F^2_{tr} +\frac{(F_{\varphi z} - C\,F_{tz})^2}{D^2}\Bigr]\,,
\label{fres1}\\[7mm]
&~& \hspace*{-1.5cm} (\partial_z \phi)^2 + e^{-2\phi}\,(\partial_z a)^2 +
3\,(\partial_z \sigma)^2 + 3\,e^{-2\sigma}\,(\partial_z b)^2=
4\Omega^2 - 2\,\Bigl[F^2_{tr} +\frac{(F_{\varphi z} - 
C\,F_{tz})^2}{D^2}\Bigr]\,.
\label{fres2}
\een
For all the solutions found above, the right-hand side of the
above equations is a constant and, as a result, the dependence of the
scalar fields on the coordinates $r$ and $z$ can be linearly at the most.
When one considers the contribution of the electromagnetic field in the
right-hand-side of eqs.~(\ref{fres1})-(\ref{fres2}), it turns out that
these solutions form two distinct groups~: for the Som-Raychaudhuri
(\ref{som}) and the Rebou\c{c}as (\ref{reb}) ansatzes, the contribution
is identically zero while, for the Som-Raychaudhuri-dual (\ref{somdual})
and the Rebou\c{c}as-dual (\ref{rebdual}) ansatzes, is not zero. This means
that, for the first group, the
scalar fields depend linearly only on the coordinate $z$ while, 
for the second group, a linear dependence also on the radial coordinate 
seems to be possible. However, when the corresponding expressions for 
the scalar fields are substituted into their equations of motion, the
radial dependence disappears and the final solution takes the form
\be
\phi=\phi_0+\phi_1\,z \quad , \quad a=a_0 \quad , \quad \label{solgr1}
\sigma=\sigma_0+\sigma_1\,z \quad , \quad b=b_0\,\,,
\label{scalar}
\ee
where now
\be
\phi_1^2 + 3\,\sigma_1^2=4\Omega^2\,\,.
\ee
The above solution strongly resembles the one derived in the neutral case.
This result was to be expected due to the absence of any interaction between
the electromagnetic and the scalar fields. We may then conclude that the only
effect of the introduction of the electromagnetic field is the modification
of the relation between the fundamental parameters of the theory~: instead
of $4\Omega^2=m^2$ holding in the neutral case, we now have
\be
4\Omega^2-m^2=Q^2\,\,,
\label{fundam}
\ee
where $Q$ plays the role of the electric or magnetic charge depending on the
``flavour" of the component of the electromagnetic field considered each
time. A result similar to (\ref{scalar}) for a single scalar field minimally
coupled to gravity and in the presence of an electromagnetic field was
derived in \cite{reb}. Here, we presented the result for four scalar fields
and for all possible configurations of the electromagnetic
field that lead to a G\"{o}del-type universe. Once again, the generalization
of the above result for the case of multiple, minimally-coupled to the
gravity, scalar fields is straightforward.
\paragraph{}
Finally, the question concerning the existence or not of Closed Timelike
Curves in the presence of the electromagnetic field has to be addressed.
The new relation (\ref{fundam}) between the fundamental parameters of the
theory leads to the result $4\Omega^2 > m^2$. When this inequality is
put together with the expression (\ref{ctcs}), we may easily conclude that 
the appearance of Closed Timelike Curves for a certain range of values
of the radial coordinate $r$ is inevitable. No matter how harmless 
the modification of the relation between the parameters of the theory
may seem at first sight, it causes the breakdown of the causality in
our G\"{o}del-type axially-symmetric universe (\ref{godel}).


\sect{EM~: Non-minimal Coupling}
\paragraph{}
In the previous section, we considered all the possible configurations
of an electromagnetic field that may lead to a charged G\"{o}del-type universe.
Although all of the solutions found are theoretically acceptable since they
constitute solutions of the equations of motion, some of the ansatzes
concerning the non-vanishing components of the electromagnetic field
are more physically preferable. If we accept the assumption of a homogeneous
distribution of electric charges over the universe, then, in the case
of a G\"{o}del-type universe, every point-charge rotating around the
$z$-axis creates a current loop. The net effect of all these tiny currents
is the creation of multiple  ``solenoids" around the $z$-axis with radii
varying from zero to infinity. An observer located in a point along the
$z$-axis would detect a constant magnetic field induced by the total current
carrying by the ``solenoids". 

So, if we assume that $\vec{B}=\hat{z}\,\mu_0\,i$, where $i$ the total
current and $\mu_0$ the permeability of the free space, and go back to
constraints (\ref{gcon1})-(\ref{gcon3}), we observe that, although the
conditions (\ref{cons}) for a G\"{o}del-type spacetime are still fulfilled,
the constraint (\ref{gcon2}) is not satisfied in the case of the minimal
coupling, i.e. in the case of a constant coupling function $f_3$. However,
if we assume that the electromagnetic field is non-minimally coupled to
gravity and assume the space-dependence of the coupling functions $f_3$
and $f_4$, the following relation
\be
4\Omega^2 - m^2 = \mu_0^2\,i^2
\label{fund2}
\ee
between the fundamental parameters is obtained once we assume that
$f_3(r)=-D^2(r)/4$. Note that, as in the minimal-coupling case, the
coupling function $f_3$ has to be negative in order to ensure the 
positivity of the $tt$-component of the energy-momentum tensor of the
electromagnetic field. 

When the ansatz for the electromagnetic field is substituted into Maxwell's
equations (\ref{abeq1})-(\ref{abeq4}), we obtain the following expressions
for the electric-charge density and the density of electric current along the
$\hat{\varphi}$ direction
\be
\rho= \frac{\mu_0\,i}{4}\,\frac{(C\,D)'}{D} \qquad , \qquad
j^{\varphi}=-\frac{\mu_0\,i}{4}\,\frac{D'}{D}\,\,,
\ee
respectively. While the density of the electric current vanishes on the
$z$-axis and becomes constant at infinity as expected, the electric-charge
density also vanishes at the origin but diverges at infinity. This ill
behaviour of the electric-charge density makes our ansatz considerably
less attractive. Moreover, a non-trivial source-term of the form
$j^t A_t$ enters the Lagrangian putting in risk the G\"{o}del-type character
of our spacetime.

However, the consideration of a constant, magnetic field along the $z$-axis
does not complete the picture of a charged, rotating universe. There are
additional components of the electromagnetic field which are highly probable
to exist and should be accommodated in our theory. For example, in a rotating
universe, there might be a steady distribution of electric or magnetic
monopoles along the $z$-axis, thus, creating a radial electric or magnetic
field, respectively. We start with the introduction of a radial electric
field~: the conditions (\ref{cons}) for the existence of a G\"{o}del-type
universe are fulfilled and the fundamental relation (\ref{fund2}) is
recovered, if the electric field and the coupling function $f_3$
satisfy the following relations
\be
F_{tr}=\frac{F_{\varphi r}}{C(r)}=\frac{\mu_0\,i}{C^2(r)} \qquad , \qquad
f_3=-\frac{C^2(r)}{4}\,\,,
\label{dem}
\ee
respectively. The substitution of the combination of $F_{\varphi r}$ and
$F_{tr}$ into Maxwell's equations (\ref{abeq1})-(\ref{abeq4}) leads to
the elimination of the charge-current density $j^{\varphi}$ while preserves
the existence of the divergent electric-charge current $\rho$ and the
corresponding source-term $j^t A_t$ in the Lagrangian.

As a last resort, we include in the theory a radial magnetic field
$F_{z\varphi}$. By following a similar procedure, a G\"{o}del-type universe
does indeed arise if we assume, apart from the relations (\ref{dem}), that 
$F_{z\varphi}=D(r) F_{tr}$. When the three components $F_{\varphi r}$,
$F_{tr}$ and $F_{z\varphi}$ are substituted into Maxwell's equations,
we see that the above combination of the components of the electromagnetic
field satisfies the source-free Maxwell's equations and that the ill-behaved
electric-charge density $\rho$ is eliminated once we assume that
$f_4=f_3$. Actually, the above combination of electromagnetic components
is a Rebou\c{c}as-dual type of solution which in the minimal-coupling case
satisfied the source-free equations as well. In the present case, 
we demanded a constant, magnetic field along the $z$-axis and, subsequently,
a radial electric and magnetic field according to the expected distribution
of charges and currents in a rotating universe. Then, the introduction of
space-dependent coupling functions ensured the existence of a
G\"{o}del-type rotating universe. 

Although the magnetic field takes on a constant value as we assumed
motivated by the corresponding behaviour of a magnetic field in Minkowski
spacetime, the other two components adopt a different behaviour. For a
linear, constant distribution of electric or magnetic charges, in Minkowski
spacetime, we anticipate to find a radial electric or magnetic field
decreasing with the radial distance from the origin. However, in a 
G\"{o}del-type spacetime, the electric field behaves as $E_r \sim 1/r^2$ near
the origin while decreases exponentially at infinity. On the other hand,
the radial magnetic field scales as $B_r \sim 1/r$ near the origin, as
expected, but assumes a constant value at infinity which leads to an
infinite total amount of magnetic charge. At the end of the day, the
corresponding magnetic-charge density associated with $B_r$ cancels out the
divergent electric-charge density $\rho$ leaving the three electromagnetic
components to satisfy the source-free Maxwell's equations.

Finally, let us add here that the fundamental relation (\ref{fund2}) leads
again to $4\Omega^2 > m^2$ and to the existence of Closed Timelike Curves
according to eq.~(\ref{ctcs}). It seems that the introduction of an
electromagnetic field, coupling minimally or non-minimally to gravity
and being induced by an electric or magnetic charge or electric current,
always results in the breakdown of causality in a G\"{o}del-type universe.
The accommodation of multiple, minimally-coupled to gravity, scalar fields
of the form (\ref{scalar}) although compatible with the above picture can
not lead to the restoration of causality of our G\"{o}del-type universe.


\sect{EM and Scalar Fields~: String Coupling}
\paragraph{}
In this section, we assume the presence of an electromagnetic and scalar
fields minimally-coupled to gravity but with a non-trivial interaction
between them. Thus, we assume that the coupling functions $f_3$
and $f_4$ are field-dependent (\ref{string}) in agreement with the one-loop
corrected superstring effective theory. For simplicity, we are going to
consider the case where only the dilaton $\phi$ and axion $a$ fields are
present in the theory and ignore the moduli fields $b$ and $\sigma$. More
specifically, we consider the following action functional
\be
S_{eff} = \int d^4x\,\sqrt{-g}\,\Biggl\{\frac{R}{2}-\Lambda  
+ \frac{1}{4}\,(\partial_\mu \phi)^2 
+ \frac{1}{4}\,e^{-2 \phi}\,(\partial_\mu a)^2 
-\frac{\alpha' e^\phi}{8g^2}\,F^{\mu \nu} F_{\mu \nu}
-\frac{\alpha' a}{8g^2}\,F {\tilde{F}}\Biggr\}\,\,.
\ee 

The possible configurations of the electromagnetic field capable of leading
to G\"{o}del-type cosmological solutions, in the minimal-coupling case, were
studied in section 4. Here, we will attempt to generalize those ansatzes
in the case of field-dependent coupling functions in such a way as to
ensure the existence of a G\"{o}del-type universe. We anticipate that the 
dependence of the coupling functions on the scalar fields will be 
appropriately absorbed in the ansatzes for the electromagnetic components
as well as in the expressions of the electric $j^\mu$ and magnetic $k^\mu$
four-currents. However, the coupling between the electromagnetic and scalar
fields may result in a spacetime dependence for the fields $\phi$ and $a$
different from the one found in section 4. We may, thus,  write the
expressions for the scalar fields in the following way
\be
\phi(r, z)=\phi^{(0)}(z) + \alpha'\,\tilde{\phi}(r,z) \quad , \quad
a(r, z)=a^{(0)} + \alpha'\,\tilde{a}(r,z)\,\,,
\label{full}
\ee
where $\phi^{(0)}$ and $a^{(0)}$ are the zero-order solution for the scalar
fields which follow if we set $\alpha'=0$ and which are given by
eq.~(\ref{solpa0}). Since, in the above Lagrangian, we have considered only
terms up to ${\cal O}(\alpha')$ in the Regge slope, we must ignore all terms
in the equations of motion that are of ${\cal O}(\alpha'^2)$ or higher. As
a result, the coupling functions $f_3$ and $f_4$ which are already of
${\cal O}(\alpha')$ will contain only the zero-order expression for the
scalar fields and have the form
\be
f_3= -\frac{\alpha' e^{\phi_0+\phi_1 z}}{8g^2} \qquad , \qquad
f_4=-\frac{\alpha' a_0}{8g^2}\,\,.
\ee

We start with the Som-Raychaudhuri generalized ansatz for a magnetic field
which can be written as
\be
F_{\varphi r}(r, z)=Q\,e^{-\phi/2}\,D(r) \,\,.
\label{dual1}
\ee
If we allow the above magnetic field to satisfy the
generalized Maxwell's equations (\ref{max}), we are led to the determination
of the corresponding electric and magnetic-charge density necessary to
support this magnetic field. They are given by
\be
\rho=\frac{\alpha' \Omega}{4g^2}\,Q\,e^{\phi/2} \qquad ,
\qquad \eta= \frac{\alpha' \phi_1}{16g^2}\,Q\,e^{-\phi/2}\,\,,
\ee
where, now, from eq.~(\ref{fund0}) $\phi_1^2=m^2$ and where we have set
$a_0=1$ for simplicity. The more complex Rebou\c{c}as ansatz now takes the form
\be
F_{tz}=Q\,e^{-\phi/2}\,sin(2\Omega z) \quad , \quad 
F_{\varphi z}=C(r)\,F_{tz} \quad , \quad
F_{\varphi r}=-Q\,e^{-\phi/2}\,D(r)\,cos(2\Omega z)
\label{nonreb}
\ee
and the corresponding electric and magnetic-charge density are found to be
\be
\rho=\frac{\alpha' \phi_1}{16g^2}\,Q\,e^{\phi/2} sin(2\Omega z)
\qquad , \qquad 
\eta= -\frac{\alpha' \phi_1}{16g^2}\,Q\,e^{-\phi/2} cos(2\Omega z)\,\,.
\ee
The ``dual" forms of the above ansatzes demand a rather complex distribution
of electric and magnetic charges and currents. Namely, for the
Som-Raychaudhuri-dual ansatz
\be
F_{z \varphi}=Q\,e^{-\phi/2}\,D(r)\,\,,
\label{dual2}
\ee
we obtain
\ben
\rho=-\frac{\alpha' \phi_1}{16g^2}\,Q\,e^{\phi/2}\,\frac{C}{D}
\qquad &,& \qquad
\eta= -\frac{\alpha'}{8g^2}\,Q\,e^{-\phi/2}\,\frac{D'}{D}\\[5mm]
j^\varphi= \frac{\alpha' \phi_1}{16g^2}\,Q\,
\frac{e^{\phi/2}}{D}\qquad &,& \qquad k^\varphi=0\,\,,
\een
while, for the generalized Rebou\c{c}as-dual ansatz
\be
F_{tr}=\frac{Q}{\sqrt{2}}\,e^{-\phi/2} \quad , \quad 
F_{\varphi r}=C(r)\,F_{tr} \quad , \quad
F_{\varphi z}=-D(r)\,F_{tr}\,\,,
\label{nonrebdual}
\ee
we are led to the expressions 
\ben
\hspace*{-0.5cm} \rho=\frac{\alpha'Q\,e^{\phi/2}}{8\sqrt{2}g^2}\,
\Bigl(\frac{D'}{D}-\frac{\phi_1 C}{2D}\Bigr)
\quad &,& \quad
\eta= -\frac{\alpha'Q\,e^{-\phi/2}}{8\sqrt{2}g^2}\,
\Bigl(\frac{D'}{D}-\frac{\phi_1 C}{2D}\Bigr)\\[5mm]
\hspace*{-0.5cm} j^\varphi= \frac{\alpha' \phi_1 Q}{16\sqrt{2}g^2}\,
\frac{e^{\phi/2}}{D} \quad &,& \quad
k^\varphi= -\frac{\alpha' \phi_1 Q}{16\sqrt{2}g^2}\,
\frac{e^{-\phi/2}}{D}\,\,.
\een
Finally, and in order to complete the picture, we reconsider the ansatzes
(\ref{trial}) and (\ref{trialdual}), which in the minimal-coupling case were
ruled out. In its appropriately generalized form, the ansatz (\ref{trial})
reads
\be
F_{tr}=Q\,e^{-\phi/2} \qquad , \qquad F_{\varphi r}=C(r)\,F_{tr}
\label{dual3}
\ee
leading to the results
\ben
\rho=\frac{\alpha'}{8g^2}\,Q\,e^{\phi/2}\,\frac{D'}{D}
\qquad &,& \qquad
\eta= \frac{\alpha' \phi_1}{16g^2}\,Q\,e^{-\phi/2}\,\frac{C}{D}\\[4mm]
j^\varphi=0 \qquad &,& \qquad k^\varphi= -\frac{\alpha' \phi_1}{16g^2}\,Q\,
\frac{e^{-\phi/2}}{D}\,\,,
\een
while, for its ``dual" ansatz
\be
F_{tz}=Q\,e^{-\phi/2} \qquad , \qquad F_{\varphi z}=C(r)\,F_{tz}\,\,,
\label{dual4}
\ee
we obtain
\be
\rho=\frac{\alpha' \phi_1}{16g^2}\,Q\,e^{\phi/2}
\qquad , \qquad
\eta= \frac{\alpha' \Omega}{4 g^2}\,Q\,e^{-\phi/2}\,\,.
\ee

In this form, with the electric and magnetic densities of charges and
currents explicitly written, we may easily detect the {\it actual
electric-magnetic duality relations} that hold between some of the
ansatzes and connect $F_{\mu\nu}$ with 
$\tilde{F}^{\mu\nu}~=~\frac{1}{2}\,\eta^{\mu\nu\rho\sigma} F_{\rho\sigma}$.
The Som-Raychaudhuri ansatz (\ref{dual1}) and the ansatz (\ref{dual4})
are connected by the duality transformations
\be
F \rightarrow -\tilde{F} \quad, \quad j^\mu \rightarrow k^\mu \quad ,
\quad k^\mu \rightarrow -j^\mu \quad , \quad \phi \rightarrow -\phi
\ee
and the same holds for the Som-Raychaudhuri-dual ansatz (\ref{dual2}) and
the ansatz (\ref{dual3}). The dilaton participates in the duality 
transformations due to the non-trivial coupling between this field and
the electromagnetic one. The same duality but without the participation
of the dilaton field was also valid in the minimal-coupling case studied
in section 4. This duality is always valid in the source-free case,
breaks down when the electric four-current $j^\mu$ is included in the
theory and is restored only after the introduction of the magnetic
four-current $k^\mu$~\cite{dirac}. 

All of the above generalized ansatzes for the electromagnetic field
respect the conditions (\ref{cons}) and, in principle, lead to a
G\"{o}del-type rotating universe of the form (\ref{godel}). However, in
almost all of the cases, a non-trivial source-term $(j^\mu+k^\mu)A_\mu$
enters the Lagrangian and subsequently the Einstein's equations destroying
eventually the G\"{o}del-type solution for the spacetime. In the case of the
Som-Raychaudhuri-dual ansatz (\ref{dual2}), we found the non-vanishing
source-term $j^\varphi A_\varphi$ to be a constant. Such a source-term
could be absorbed in the cosmological constant of the theory leading once
again to a G\"{o}del-type universe. However, it turns out that the combination
of the source-term and the kinetic term $F_{\mu\nu} F^{\mu\nu}$ vanishes
identically which leads us to the results of section 3 for an uncharged
G\"{o}del-type universe. 

The only ansatz, for which this source-term is identically zero, is the
generalized Som-Raychaudhuri ansatz (\ref{dual1}). As a result, this is
the only acceptable solution for the electromagnetic field since it leaves
unchanged the Einstein's equations while satisfies the source-full Maxwell's
equations at the same time. On the other hand, the required distribution
of electric and magnetic charges is finite and homogeneous in radial space
and, thus, more physically favoured with the corresponding current densities
being identically zero. For this solution, the constraint (\ref{gcon2}) leads
to the following relation of the fundamental parameters of the theory
\be
4\Omega^2-m^2=\frac{\alpha'}{2g^2}\,Q^2\,\,.
\label{fund3}
\ee
We observe that the introduction of a non-trivial interaction between
the scalar fields of the theory and the electromagnetic field does not
restore the causality of the axially-symmetric spacetime since the above 
relation results, once again, in the appearance of Closed Timelike Curves.
\paragraph{}

Finally, we need to determine the solution for the scalar fields $\phi$
and $a$ in the presence of the $\alpha'$-order electromagnetic field. For
the only acceptable configuration, i.e. 
the Som-Raychaudhuri generalized ansatz (\ref{dual1}) for a magnetic
field along the $z$-axis , eqs.~(\ref{geq8}), (\ref{gres1}) and
(\ref{gres2}) take the form 
\ben
&~& \partial_r \phi\,\partial_z \phi+ e^{-2\phi}\,\partial_r a\,
\partial_z a=0\,\,, \label{soma1} \\[4mm]
&~& (\partial_r \phi)^2 + e^{-2\phi}\,(\partial_r a)^2=0\,\,,
\label{soma2}\\[4mm]
&~& (\partial_z \phi)^2 + e^{-2\phi}\,(\partial_z a)^2=4\Omega^2\,\,.
\label{soma3}
\een
By taking into account the expressions (\ref{full}), we conclude that each
of the above terms involving derivatives of the axion field $a$ as well
as the term $(\partial_r \phi)^2$ are of ${\cal O}(\alpha'^2)$ and they
should be ignored. Then, eq.~(\ref{soma2}) is trivially satisfied while
eq.~(\ref{soma1}) leads once again to the independence of the dilaton
field of the radial coordinate $r$.
Eq.~(\ref{soma3}) reveals the fact that the dilaton field is linearly
dependent on the coordinate $z$ leading to the result
\be
\phi(z) = \phi_0 + \phi_1\,z\,\Bigl(1+\frac{\alpha' Q^2}{4g^2m}\Bigr)\,\,.
\label{finalphi}
\ee

Since the Einstein's equations can give us no information on the form of 
the axion field, we turn to the corresponding equation of motion
(\ref{axion1}). The combination $F \tilde{F}$ which acts as a source-term
in the axion equation of motion is identically zero for the ansatz
(\ref{dual1}) and, thus, we obtain 
\be
e^{\textstyle -2(\phi_0+\phi_1 z)}\,\frac{\partial}{\partial r}\,\Bigl[D(r)\,
\frac{\partial\,\tilde a (r, z)}{\partial r}\Bigl]
+ D(r)\,\frac{\partial}{\partial z}\Bigl[e^{\textstyle -2(\phi_0+\phi_1 z)}\,
\frac{\partial\,\tilde a (r, z)}{\partial z}\Bigr]=0\,\,.
\ee
By applying the method of separation of variables, the solution for the
axion field takes the form
\be
a(r, z)=a_0 + \alpha' A_0\,cosh(mr)\,cos(mz)\,\,,
\ee
where $A_0$ is an arbitrary constant and where the reality of the
axion field has been demanded.

It is worth noting that the existence of a G\"{o}del-type rotating universe,
in the presence of interacting scalar and electromagnetic fields, is
incompatible with a universe empty from electric and magnetic charges and
currents. For instance, if we demand that the Som-Raychaudhuri ansatz
(\ref{dual1}) satisfies the source-free Maxwell's equations, we are led
to the result $C=0$ and
\be
D(r)=\frac{1}{\mu}\,sin(\mu r) \qquad {\rm with} \qquad 
2\Lambda \equiv \mu^2=\frac{\alpha'}{2g^2}\,Q^2\,\,.
\label{chres1}
\ee
Then, the line-element of the spacetime assumes the spherically-symmetric
form
\be
ds^2=dt^2-dr^2-dz^2-\frac{sin^2(\mu r)}{\mu^2}\,d\varphi^2
\ee
with the metric component $g_{\varphi\varphi}=-D^2(r)$ being always negative
and, thus, excluding the presence of Closed Timelike Curves. Near the origin
$D(r) \sim r$ and the above line-element resembles the Minkowski one.
Finally, from eqs.~(\ref{gres1}) and (\ref{gres2}) we find that the only
acceptable solution for the scalar fields is $\phi=\phi_0=const.$ and
$a=a_0=const.$ leading to their decoupling from the theory. The presence
or not of the source terms in Maxwell's equations affects the form of the
metric function $C(r)$ and thus the structure of the spacetime~: while in
the source-free case, the system of equations accepts only
spherically-symmetric line-elements with no CTC's and a positive cosmological
constant, in the source-full case, axially-symmetric G\"{o}del-type
solutions, that are characterized by a negative cosmological constant
and the presence of CTC's, arise. Moreover, in the former case, the scalar
fields are decoupled from the gravito-magnetic system, while, in the latter
case, a first-order, in $\alpha'$, solution for the scalar fields was found
that completes the zero-order solution found in section 3.

Let us, finally, note that, in the neutral case studied in section 3,
the negative sign of the
cosmological constant was dictated by the relation $C'^2=-2\Lambda\,D^2$
and the reality of the metric functions. In all the charged cases, however,
considered in this article, there was no constraint on the sign of 
$\Lambda$. Due to the fact that the original G\"{o}del solutions were
derived for a negative cosmological constant, we chose $\Lambda$ to have
a negative sign in these cases, too. We would like to stress that 
G\"{o}del-type solutions arise also for a positive cosmological constant.
In this case, the conditions (\ref{cons}) take the form
\be
\frac{D''}{D}=const.=-2\Lambda=-\mu^2 \qquad , \qquad
\frac{C'}{D}=const.= 2\Omega
\ee
and are still satisfied by all the configurations of the electromagnetic
field considered in sections 4, 5 and 6. However, the solution for the
spacetime line-element, now, has the form
\be
ds^2=\biggl[dt+\frac{4\Omega}{\mu^2}\,sin^2\bigl(\frac{\mu r}{2}\bigr)
\,d\varphi\biggr]^2 - dr^2-dz^2-\frac{sin^2(\mu r)}{\mu^2}\,d\varphi^2
\label{godel2}
\ee
and the relations (\ref{fundam}), (\ref{fund2}) and (\ref{fund3}) between
the fundamental parameters of the theory are modified with $m^2$ being
replaced by $-\mu^2$. The $g_{\varphi\varphi}$ component of the G\"{o}del
metric (\ref{godel2}) is given by
\be
g_{\varphi\varphi}=-\frac{4}{\mu^2}\,sin^2(\frac{\mu r}{2})\,\biggl[1-
\biggl(1+\frac{4\Omega^2}{\mu^2}\biggr)\,sin^2\bigl(\frac{\mu r}{2}\bigr)
\biggr]
\label{ctcs2}
\ee
and it is easy to see that the sign of the expression inside the brackets
changes from positive to negative and vice versa in a harmonic way giving
rise to successive causal and acausal regions in spacetime. This behaviour
was originally found in the case of a, minimally-coupled to gravity,
electromagnetic field~\cite{reboucas} and, subsequently, in the framework
of other theories as well~\cite{barrow}.


\sect{Conclusions}
\paragraph{}
In this paper, we have addressed the question of the existence of
axially-symmetric G\"{o}del-type solutions in the framework of gravitational
theories formulated in a way similar to the one-loop superstring effective
theory. The incorporation of minimally-coupled to gra\-vity scalar fields as
well as a non-vanishing electromagnetic field, either minimally
or non-minimally coupled to gravity, was proved to be consistent with
the concept of G\"{o}del-type rotating universes. The coupling functions
of the electromagnetic field were assumed to be zero, constant,
space-dependent or field-dependent with various consequences each time
on the structure of spacetime and the solution for the scalar fields.

In the absence of an electromagnetic field, the system of the four scalar
fields of the superstring effective theory, all minimally-coupled to gravity,
was studied in an axially-symmetric background. Assuming also a vanishing
cosmological constant, the above system inevitably leads to a
spherically-symmetric spacetime background and to the decoupling of the
scalar fields from the theory. Once a non-vanishing, negative value for the
cosmological constant is assumed, an axially-symmetric solution for the
spacetime of a G\"{o}del type does arise with the scalar fields either
depending linearly on the $z$ coordinate or adopting a constant value.
The resulting solution for the spacetime is
an ``extreme" G\"{o}del-type universe with the fundamental parameters of the
theory satisfying the relation~: $4\Omega^2=m^2$. It is exactly this 
relation that ensures the non-existence of Closed Timelike Curves 
and the non-violation of causality in our G\"{o}del-type universe. However,
as it is known in the literature, this spacetime solution is ``on the
verge" of displaying the breakdown of causality which is realized when
$4\Omega^2$ becomes even slightly larger that $m^2$. The introduction
of multiple, minimally-coupled to gravity, scalar fields is straightforward
leaving the solution for the spacetime unchanged.

Next, we studied the charged case, and the corresponding equations of motion
were re-written in a convenient way which allowed us to determine first the
solution for the metric functions in terms of the electromagnetic ansatz
considered each time and subsequently the solution for the scalar fields.
We started with the case of minimally-coupled electromagnetic and scalar
fields assuming constant values for the two coupling functions $f_3$ and
$f_4$. The gravito-electromagnetic system was examined thoroughly, in 
the presence of a negative cosmological constant, and
apart from the two solutions already known in the literature, the 
Som-Raychaudhuri solution and the Rebou\c{c}as solution, two new configurations
were determined. They follow from the aforementioned solutions once one
interchanges the indices $r$ and $z$ in the
non-vanishing components of the electromagnetic field. The invariance of
the equations of motion under this interchange results in the fulfillment
of the G\"{o}del conditions by these new ansatzes as well. As a result, a
G\"{o}del-type rotating universe emerges but, now, the relation between
the fundamental parameters of the theory has become $4\Omega^2>m^2$.
This means that Closed Timelike Curves appear for a certain range of
the radial coordinate and the causality is violated. The introduction
of the electromagnetic field has evidently destroyed the fragile balance
between the violation and the conservation of causality that characterized
the neutral case. On the other hand, the scalar fields assume a form similar
to the one found in the neutral case due to the absence of any interaction
with the electromagnetic field.

The more physically favoured ansatz of a constant magnetic field along the 
rotation axis and a radial electric and magnetic component was studied next.
This specific ansatz follows from the anticipated distribution of magnetic
and electric charges and currents in a rota\-ting universe~: point-charges
that create current loops around the rotation axis and a static distribution
of electric and magnetic monopoles along the same axis. For such a 
configuration, the G\"{o}del conditions are again fulfilled and the Maxwell's
equations are satisfied as long as we assume that both of the coupling
functions $f_3$ and $f_4$ have the same dependence on the radial coordinate. 
Then, a G\"{o}del-type solution for the spacetime does arise which is again
characterized by the violation of causality and the appearance of Closed
Timelike Curves. The incorporation of minimally-coupled scalar fields
although straightforward and compatible with the above results can not
restore the causality in the universe.

Finally, the coupling functions $f_3$ and $f_4$ were assumed to have the 
field-dependent form of the superstring effective theory. In this case,
there is a non-trivial coupling between the scalar fields of the theory
and the electromagnetic field. The possible configurations of the
electromagnetic field that could lead to a G\"{o}del-type rotating universe
were re-examined. Several such configurations were found but only one
of them, the generalized Som-Raychaudhuri ansatz of a magnetic field
along the rotation axis, managed to satisfy the G\"{o}del conditions and the
corresponding source-full Maxwell's equations at the same time. Once
again, the relation $4\Omega^2>m^2$, that holds in this case too, leads
to the appearance of  Closed Timelike Curves. The non-trivial coupling
between the scalar fields and the electromagnetic field, although not
able to restore the causality, leads to different results for the scalar
fields and especially for the axion field $a$. While the dilaton field
depends only linearly on the $z$-coordinate as in the previous cases,
the axion field abandons the constant value adopted so far and assumes
a form which strongly depends on both $r$ and $z$ coordinates.

In conclusion, a charged, string-inspired, gravitational theory of
scalar and electromagnetic fields of the form considered in this article
always accepts axially-symmetric G\"{o}del-type solutions in the presence
of a non-vanishing, negative cosmological constant. G\"{o}del-type
solutions arise also in the charged case with a positive cosmological
constant, however, they are shown to lead to successive causal and acausal
regions in spacetime. Only in the neutral case, where we ignore the
electromagnetic field, we are able to ensure the non-violation of causality
in the universe in agreement with previous results by Barrow and
Dabrowski~\cite{barrow}. Once the electromagnetic field is introduced,
the emergence of the G\"{o}del-type solution is always accompanied by
the appearance of Closed Timelike Curves and the violation of causality.
However, the realization of the causally-ill G\"{o}del solutions in the
universe should be impeded by physical reasons. The uniformity of the
Cosmic Microwave Background Radiation, as dictated by the COBE data,
leads to the conclusion that the rate of the rigid rotation of our
universe, if existent at all, should be extremely small. On the other 
hand, there is only weak indication for a global charge in the
universe with the bulk of astrophysical objects being neutral. As
a result, it is highly improbable that our universe would adopt the
geometry of a G\"{o}del cosmological solution, thus, excluding the
possibility of the appearance of Closed Timelike Curves and the
violation of causality. Finally, an interesting question that still
remains open is what is the effect of the higher-order perturbative
terms in $\alpha'$ on the causal pathologies of the G\"{o}del-type
solutions derived in the framework of the superstring effective action.
It is rather plausible that the presence of these terms, even if it
is not capable of resolving the appearance of CTC's, would render the
stringy G\"{o}del-type solutions unstable, thus, depriving them from
any physical significance.
\\[5mm]
{\bf Acknowledgments} \,P.K. would like to thank Marcelo J. Rebou\c{c}as
for enlightening discussions. P.K. would also like to acknowledge
financial support from the research program $\Pi$ENE$\Delta$-95
of the Greek Ministry of Science and Technology and from DOE grant
DE-FG02-94ER40823 at Minnesota during the early and late stages,
respectively, of this work.

\def\theequation{A.\arabic{equation}}
\def\thesection{Appendix:}

\section{Useful formulae and equations}
\paragraph{}
For the axially-symmetric ansatz (\ref{metric}) for the metric tensor
of the universe, we find the following results for the non-vanishing
components of the fully-covariant curvature tensor $R_{\mu\nu\rho\sigma}$ 
\ben
&~& R_{trtr} = \frac{C'^2}{4D^2}  \qquad , \qquad 
R_{trt\varphi} = \frac{C {C'}^2}{4D^2}\,\,,\\[5mm]
&~& R_{trr\varphi} = -\frac{C''}{2} -\frac{C {C'}^2}{4D^2} +\frac{C'D'}{2D}
\qquad , \qquad R_{t\varphi t \varphi} = \frac{{C'}^2}{4}\,\,,\\[5mm]
&~& R_{r\varphi r\varphi} =C\,C''-D D''+\frac{3}{4}\,{C'}^2 +
\frac{C {C'}^2}{4D^2}-\frac{C C'D'}{D}\,\,,
\een
the Ricci tensor $R_{\mu\nu}$
\ben
&~& R_{tt} = -\frac{{C'}^2}{2D^2} \quad , \quad
R_{rr} = \frac{D''}{D} -\frac{{C'}^2}{2D^2}\,\,,\\[5mm]
&~& R_{\varphi\varphi} = -C\,C''+D D''-\frac{{C'}^2}{2}-
\frac{C^2 {C'}^2}{2D^2}+\frac{C C'D'}{D}\,\,,\\[5mm]
&~& R_{t\varphi}= -\frac{C''}{2}-\frac{C'}{2D^2}\,(C C'-D D')
\een
and the scalar curvature $R$
\be
R=-\frac{2D''}{D} + \frac{{C'}^2}{2D^2}\,\,,
\ee
respectively. By using the above expressions, the non-vanishing
components of the Einstein tensor $G_{\mu\nu}$ are the following
\ben
&~& G_{tt}=-\frac{3}{4}\,\frac{{C'}^2}{D^2}+\frac{D''}{D} \qquad ,
\qquad G_{rr}=-\frac{{C'}^2}{4D^2}\,\,, \\[4mm]
&~& G_{\varphi\varphi}=-C\,C''+\frac{C^2 D''}{D}-\frac{{C'}^2}{4}-
\frac{3 C^2 {C'}^2}{4D^2}+\frac{C C'D'}{D}\,\,,\\[4mm]
&~& G_{zz}=\frac{{C'}^2}{4 D^2}-\frac{D''}{D}\,\,,\\[4mm]
&~& G_{t\varphi}=-\frac{C''}{2}+\frac{C D''}{D}-
\frac{3 C {C'}^2}{4D^2}+\frac{C'D'}{2 D}\,\,.
\een

If we assume that, due to the symmetry of the line-element of the spacetime,
the scalar fields depend only on $r$ and $z$, the corresponding components
of the energy-momentum tensor $T_{\mu\nu}^s$ take the form
\ben
&~& T_{tt}^s=\frac{1}{4}\,\biggl\{\,(\partial_r \phi)^2 + (\partial_z \phi)^2 +
e^{-2\phi}\,[\,(\partial_r a)^2 + (\partial_z a)^2\,] +
3\,[\,(\partial_r \sigma)^2 + (\partial_z \sigma)^2\,]+\nonumber\\[3mm]
&~& \hspace*{1.8cm} 3e^{-2\sigma}\,[\,(\partial_r b)^2 +
(\partial_z b)^2\,]\,\biggr\}\,\,,\\[4mm]
&~& T_{rz}^s= \frac{1}{2}\,\biggl\{\,\partial_r \phi\,\partial_z \phi+
e^{-2\phi}\,\partial_r a\,\partial_z a+
3\,\partial_r \sigma\,\partial_z \sigma+
3e^{-2\sigma}\,\partial_r b\,\partial_z b\,\biggr\}\,\,,\\[4mm]
&~& T_{rr}^s=\frac{1}{4}\,\biggl\{\,(\partial_r \phi)^2 - (\partial_z \phi)^2
+ e^{-2\phi}\,[\,(\partial_r a)^2 - (\partial_z a)^2\,] +
3\,[\,(\partial_r \sigma)^2 - (\partial_z \sigma)^2\,]+\nonumber\\[3mm]
&~& \hspace*{1.8cm} 3e^{-2\sigma}\,[\,(\partial_r b)^2 -
(\partial_z b)^2\,]\,\biggr\}\,\,,\\[4mm]
&~&T_{zz}^s=-T_{rr}^s \quad , \quad T_{\varphi\varphi}^s=
(C^2-D^2)\,T_{tt}^s \quad , \quad T_{t\varphi}^s=C\,T_{tt}^s\,\,.
\een
\paragraph{}
In the presence of a non-vanishing electromagnetic field, the Einstein's
equations take the form
\ben
&~& \hspace*{-1cm} -\frac{3}{4}\,\frac{{C'}^2}{D^2}+\frac{D''}{D} +\Lambda +
4f_3\,\Bigl(g^{rr} F_{tr}^2+g^{zz} F_{tz}^2+
g^{\varphi\varphi} F_{t\varphi}^2 - \frac{F^2}{4}\Bigr) = -T_{tt}^s\,\,, 
\label{geq1} \\[6mm]
&~& \hspace*{-1cm} -\frac{{C'}^2}{4D^2} - \Lambda +4f_3\,
\Bigl(g^{tt} F^2_{tr}
+ g^{zz} F_{rz}^2+ g^{\varphi\varphi} F^2_{r\varphi} +
2g^{t\varphi} F_{tr} F_{\varphi r} +\frac{F^2}{4}\Bigr) =- T_{rr}^s\,\,,
\label{geq2}\\[6mm]
&~& \hspace*{-1cm} \frac{{C'}^2}{4 D^2}-\frac{D''}{D} - \Lambda + 
4f_3\,\Bigl(g^{tt} F_{zt}^2+g^{rr} F_{zr}^2+
g^{\varphi\varphi} F_{z\varphi}^2+2 g^{t\varphi} F_{zt} F_{z\varphi} +
\frac{F^2}{4}\Bigr)=-T_{zz}^s\,\,, \label{geq3}\\[6mm]
&~& \hspace*{-1cm} -C\,C''+\frac{C^2 D''}{D}-\frac{{C'}^2}{4}-
\frac{3 C^2 {C'}^2}{4D^2}+\frac{C C'D'}{D}+\Lambda\,(C^2-D^2)
\nonumber \\[4mm]
&~& \hspace*{2cm}+\,\,4 f_3\,\Bigl[ g^{tt} F_{\varphi t}^2+
g^{rr} F_{\varphi r}^2 + g^{zz} F^2_{\varphi z} -
\frac{F^2}{4}\,(C^2-D^2)\Bigr]=
-T_{\varphi\varphi}^s\,\,, \label{geq4}\\[6mm]
&~& \hspace*{-1cm} -\frac{C''}{2}+\frac{C D''}{D}-\frac{3 C {C'}^2}{4D^2}+
\frac{C'D'}{2 D} + \Lambda \,C \nonumber \\[4mm]
&~& \hspace*{2cm} +\,\,4 f_3\,\Bigl(
g^{rr} F_{tr} F_{\varphi r} + g^{zz} F_{tz} F_{\varphi z} -
g^{t\varphi} F_{t\varphi}^2 - \frac{F^2}{4}\,C\Bigr) =-T_{\varphi t}^s\,\,, 
\label{geq5}\\[7mm]
&~& \hspace*{-0.7cm} g^{zz} F_{tz} F_{rz} + g^{\varphi\varphi} F_{t \varphi}
F_{r\varphi} + g^{t\varphi} F_{t\varphi} F_{rt}=0\,\,, \label{geq6}\\[7mm]
&~& \hspace*{-0.7cm} g^{rr} F_{tr} F_{zr} + g^{\varphi\varphi} F_{t \varphi}
F_{z\varphi} + g^{t\varphi} F_{t\varphi} F_{zt}=0\,\,, \label{geq7}\\[7mm]
&~& \hspace*{-0.7cm} 4f_3\,\Bigl[g^{tt} F_{rt} F_{zt} +
g^{\varphi\varphi} F_{r \varphi} F_{z\varphi} +
g^{t\varphi}(F_{rt} F_{z\varphi} + F_{zt} F_{r\varphi})\Bigr]=
-T^s_{rz}\,\,, \label{geq8}\\[7mm]
&~& \hspace*{-0.7cm} g^{tt} F_{rt} F_{\varphi t} + g^{zz} F_{rz}
F_{\varphi z} + g^{t\varphi} F_{r\varphi} F_{\varphi t}=0\,\,,
 \label{geq9}\\[7mm]
&~& \hspace*{-0.7cm} g^{tt} F_{zt} F_{\varphi t} + g^{rr} F_{zr}
F_{\varphi r} + g^{t\varphi} F_{z\varphi} F_{\varphi t}=0\,\,,
\label{geq10}
\een \\
where $F^2=F^{\mu\nu} F_{\mu\nu}$ is given by
\ben
F^2 &=& 2\,\Biggl\{g^{tt}\Bigl(-F^2_{tr} - F^2_{tz} + g^{\varphi\varphi}
F^2_{t\varphi}\Bigr) + F^2_{rz} - g^{\varphi\varphi}\Bigl(F^2_{r\varphi}
+ F^2_{z\varphi}\Bigr)- \nonumber\\[2mm]
&~& \hspace*{0.5cm} \Bigl(g^{t\varphi} F_{t\varphi}\Bigr)^2-
2g^{t\varphi}\Bigl(F_{tr}\,F_{\varphi r}+F_{tz} F_{\varphi z}\Bigr)
\Biggr\} 
\een
and $T_{\mu\nu}^s$ denotes the energy-momentum tensor of the scalar fields
given above. 

In the charged case, the components of Einstein's equations are supplemented
by a set of four additional equations, the Maxwell's equations, that come
from the equation of motion of the electromagnetic field (\ref{abelian})
and have the form
\vspace*{2mm}
\ben
&~& \hspace*{-0.5cm} \partial_r\biggl\{\frac{f_3}{D}\,
\Bigl[(D^2-C^2)\,F_{rt}+ C\,F_{r\varphi}\Bigr]\biggr\} +
\partial_r \Bigl(f_4\,F_{z\varphi}\Bigr)\nonumber \\[2mm] &~&  \hspace*{2cm}
+ \,\frac{1}{D}\,\,\partial_z\,\biggl\{f_3\,\Bigl[(D^2-C^2)\,F_{zt}+
C\,F_{z\varphi}\Bigr]\biggr\}-
\partial_z\,\Bigl(f_4\,F_{r\varphi}\Bigr)=0\,\,, \label{abeq1}
\\[6mm]
&~& \hspace*{-0.5cm} D\,\,\partial_z\,\Bigl(f_3\,F_{zr}\Bigr)-
\partial_z\,\Bigl(f_4\,F_{t\varphi}\Bigr)=0\,\,, \label{abeq2}\\[6mm]
&~& \hspace*{-0.5cm} \partial_r\,\Bigl(D\,f_3\,F_{rz}\Bigr)+
\partial_r\Bigl(f_4\,F_{t\varphi}\Bigr)=0\,\,, \label{abeq3}\\[6mm]
&~& \hspace*{-0.5cm} \partial_r\biggl\{\frac{f_3}{D}\,
\Bigl(F_{r\varphi}-C\,F_{rt}\Bigr)\biggr\} + \frac{1}{D}\,
\partial_z\,\biggl\{f_3\,\Bigl(F_{z\varphi}-C\,F_{zt}\Bigr)\biggr\}
\nonumber \\[2mm] &~& \hspace*{6cm}-\,\partial_r \Bigl(f_4\,F_{tz}\Bigr)+
\partial_z\,\Bigl(f_4\,F_{tr}\Bigr)=0 \,\,. \label{abeq4}
\een
As a result, any solution for the metric functions $C(r)$ and $D(r)$ and
the various components of the electromagnetic field has to satisfy also
the above four equations.


\end{document}